\documentstyle[twoside,fleqn,espcrc2]{article}

\input epsf
\def\Re{\mbox{Re}}

\def\epsp{ $\varepsilon '/\varepsilon$ }

\title{$\Delta I=1/2$ rule from staggered fermions.}

\author{D. Pekurovsky and G. Kilcup\address{Department of Physics,
the Ohio State University, 174 W. 18th Ave., Columbus OH 43210, USA}}
       
\begin{document}

\begin{abstract}
 We present our latest results for the $\Delta I=1/2$ rule,
obtained on quenched ensembles with 
$\beta =6.0$ and 6.2, and a set of $N_f = 2$ configurations with 
$\beta =5.7$. The statistical noise is quite under control.
We observe an enhancement of the $\Delta I=1/2$ amplitude 
consistent with experiment, although the systematic errors are still large. 
We also present a non-perturbative determination of $Z_P$, $Z_S$ and 
the strange quark mass. We briefly discuss our progress in 
calculating \epsp . 
\end{abstract}

\maketitle

\section{Introduction and methods}

It is well-known that the $\Delta I=1/2$ channel 
of non-leptonic kaon decays is enhanced compared to the 
$\Delta I=3/2$ one. 
In particular, $\omega \equiv \Re A_0/\Re A_2 = 22$, where
$A_{0,2}e^{i\delta_{0,2}} \equiv \langle (\pi\pi )_{I=0,2}|H_W|K^0\rangle $
This talk is a status report on our work in calculating these 
matrix elements (MEs) using staggered fermions.
%We have obtained enough 
%statistics to bring the noise down to an acceptable
%level. 
We have also computed MEs relevant for \epsp .

%Previous work??
 
The effective weak Hamiltonian for this problem is as follows:

\begin{equation}
H_W^{\it eff} = 
\frac{G_F}{\sqrt{2}} V_{ud}\,V^*_{us} \sum_{i=1}^{10} \Bigl[
z_i(\mu) + \tau y_i(\mu) \Bigr] O_i (\mu) 
 \, , \nonumber
\end{equation}
where $\tau = - V_{td}V_{ts}^{*}/V_{ud} V_{us}^{*}$, $z_i$ and $y_i$
are Wilson coefficients. We work in the standard basis of the 10 
four-fermion operators defined in~\cite{buras}. 

Our task is to compute $\langle \pi\pi |O_i|K\rangle $. 
Putting a two-pion state on the lattice is a well-known 
technical problem. We use the chiral perturbation theory  
method~\cite{bernard} to relate \mbox{$\langle \pi\pi |H_W|K\rangle $} 
to $\langle \pi |H_W|K\rangle $ and $\langle 0|H_W|K\rangle $. 
This is equivalent to subtraction of a single lower-dimension operator 
\mbox{$O_{sub}=(m_s+m_d)\overline{s}d + (m_d-m_s)\overline{s}\gamma_5d$,}
which is the only operator allowed to mix due to the chiral properties
of staggered fermions. 
This procedure does not take into account the higher
order corrections in chiral perturbation theory (ChPT), in particular 
the final state interactions of the pions. These corrections are known to be
large, which introduces a big systematic uncertainty in our results.

We follow the strategy of ME computation with staggered 
fermions~\cite{kilcup1}  and compute
three types of fermion contractions, known as ``eight'', ``eye''
and ``annihilation'' diagrams. The latter two types are quite noisy, 
but we gained enough statistics to bring the noise down to an 
acceptable level for all basic operators $O_1$--$O_{10}$.

Table~1 shows our simulation parameters. Our lattice is 
replicated 4 times in time direction to avoid contamination from
excited states. We use degenerate mesons \mbox{($m_s=m_d=m_u$)} and
gauge-invariant, tadpole-improved operators with staggered fermions.

\begin{table*}[htb]
% space before first and after last column: 1.5pc
% space between columns: 3.0pc (twice the above)
\setlength{\tabcolsep}{1.5pc}
\newlength{\digitwidth} \settowidth{\digitwidth}{\rm 0}
\catcode`?=\active \def?{\kern\digitwidth}
\caption{Simulation parameters}
\label{params}
\begin{tabular*}{\textwidth}{lllllll}
\hline
$N_f$ & $\beta $ & Size & L, fm & N \\
\hline
0 & 6.0 & $16^3\times (32\times 4)$ & 1.6 & 216 \\
0 & 6.0 & $32^3\times (64\times 2)$ & 3.2 & 26 \\
0 & 6.2 & $24^3\times (48\times 4)$ & 1.7 & 26 \\
2 & 5.7 & $16^3\times (32\times 4)$ & 1.6 & 83 \\
\hline
\end{tabular*}
\end{table*}
\begin{figure}[!htb]
\begin{center}
\leavevmode
\centerline{\epsfysize=4.5cm \epsfbox{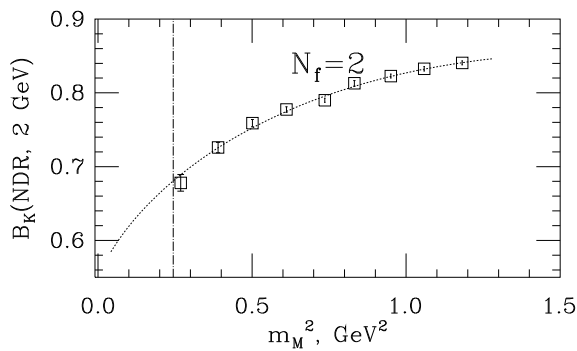}}
\vspace{-1.5cm}
\end{center}
\label{Bk}
\caption{$B_K$ vs. meson mass squared. Vertical line here and in other 
plots indicates the physical kaon mass. The fit is of the form 
$y = a + bx + cx\log{x}$. }
\vspace{-0.5cm}
\end{figure} 

\section{$\Re A_2$ and $\langle O_K\rangle $}

$\Re A_2$ can be related by ChPT to $\langle \overline{K^0}|O_K|K^0\rangle$,
so at the lowest order we just need to compute $f_K$ and $B_K$ (defined in 
$ \langle \overline{K^0}|O_K|K^0\rangle = 8/3\; m_K^2f_K^2B_K$,
where $O_K= \overline{s}\gamma_L d \;\overline{s}\gamma_L d$).
The $B_K$ parameter involves calculating only ``eight'' contractions
and is now well studied (e.g.~\cite{us,jl}). The form of the chiral 
behaviour \mbox{$B_K=a+bm_K^2+c\;m_K^2\log{m_K^2}$~\cite{sharpe1}} 
produces a reasonable fit and gives a 
finite non-zero value in the chiral limit (Fig.~1). However, 
the physical $\Re A_2$
(proportional to $\langle O_K\rangle /m_K^2 = 8/3\;f_K^2 B_K$)
is very sensitive to the meson mass, contrary to a naive expectation
(see Fig.~2). This is due to a sizeable slope in $f_K$ vs. $m_K^2$ 
(which is of the same order as the physical slope) (Fig.~3). Thus there is
a large uncertainty due to unknown higher order terms in ChPT.
\begin{figure}[!hbt]
\begin{center}
\leavevmode
%\vspace{-1cm}
\centerline{\epsfysize=5cm \epsfbox{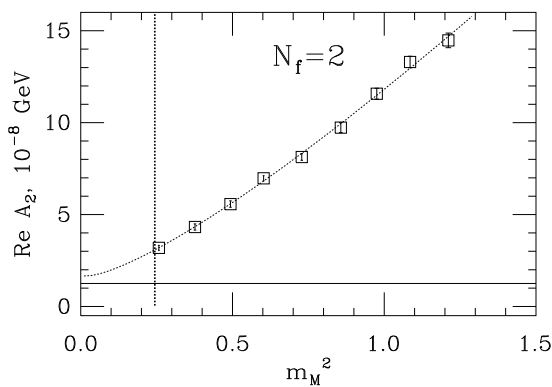}}
\vspace{-1.5cm}
\end{center}
\label{Okm}
\caption{$\Re A_2$ vs. meson mass squared. 
The fit is of the form  $y = a + bx + cx\log{x}$. The vertical line
indicates the physical kaon mass.}
\vspace{-0.5cm}
\end{figure} 
\begin{figure}[!htb]
\begin{center}
\leavevmode
\centerline{\epsfysize=4.5cm \epsfbox{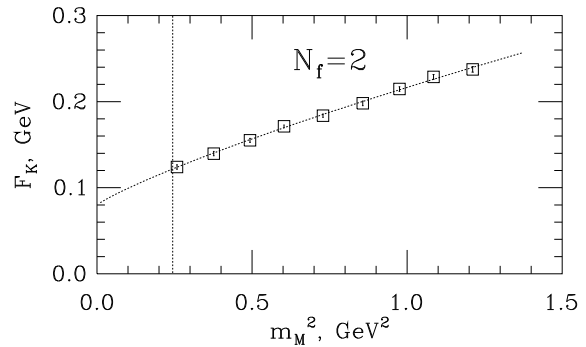}}
\vspace{-1.5cm}
\end{center}
\label{fk}
\caption{$f_K$ vs. meson mass squared. }
\vspace{-0.5cm}
\end{figure} 
Naively taking the meson mass \linebreak
\mbox{$M^2=(m_K^2+m_\pi^2)/2$} and using our quenched value 
of $B_K$ in continuum limit yields: 
$\Re A_2 = (1.67 \pm 0.02)\cdot 10^{-8}\;\mbox{GeV}$, to be 
compared with experimental $1.25\cdot 10^{-8}\;\mbox{GeV}$. 

\begin{figure}[htb]
\begin{center}
\leavevmode
%\vspace{-0.5cm}
\centerline{\epsfysize=5cm \epsfbox{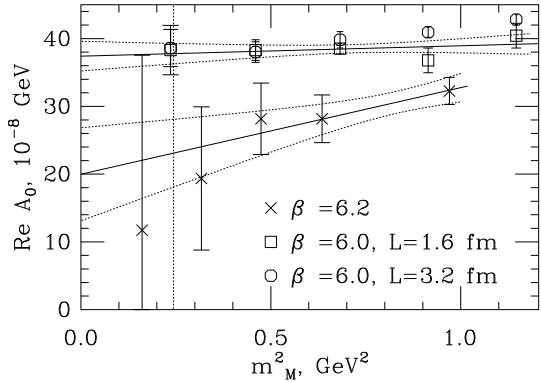}}
\vspace{-1.5cm}
\end{center}
\label{A0beta}
\caption{$\Re A_0$ vs. meson mass squared for 
quenched $\beta=6.0$ and 6.2. Finite volume study for $\beta=6.0$ 
is also shown.}
\vspace{-0.5cm}
\end{figure} 

\begin{figure}[htb]
\begin{center}
\leavevmode
\centerline{\epsfysize=5cm \epsfbox{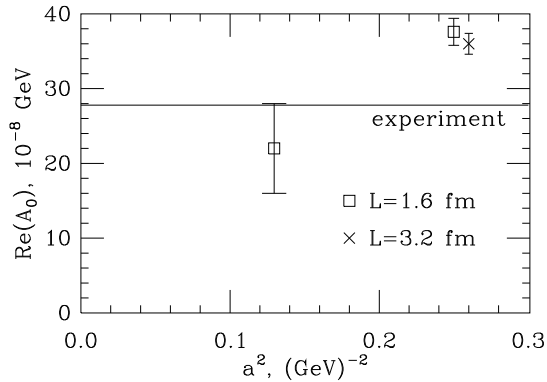}}
\vspace{-1.5cm}
\end{center}
\label{A0}
\caption{$\Re A_0$ vs. lattice spacing squared. 
The cross is the result from the larger volume $\beta=6.0$ ensemble.}
\vspace{-0.5cm}
\end{figure} 

\section{$\Re A_0$ and $\Delta I=1/2$ rule}

Results for $\Re A_0$ for quenched $\beta =6.0$ and 6.2 ensembles
are shown in Fig.~4. The dependence on the meson mass
is much smaller than that of $\Re A_2$. We have checked the lattice volume
dependence and found it to be small for lattice \mbox{sizes 1.6 fm} and above. 
However, the results significantly depend on the lattice spacing
(see Fig.~5). The $\beta =6.2$ point seems to bring the 
continuum value below the experiment. However, final state interactions 
could additionally raise it by as much as $100\%$. Finally, we checked
the effect of unquenching and found it to be small 
compared to noise (see Fig.~6).

\begin{figure}[htb]
\begin{center}
\leavevmode
\centerline{\epsfysize=4.5cm \epsfbox{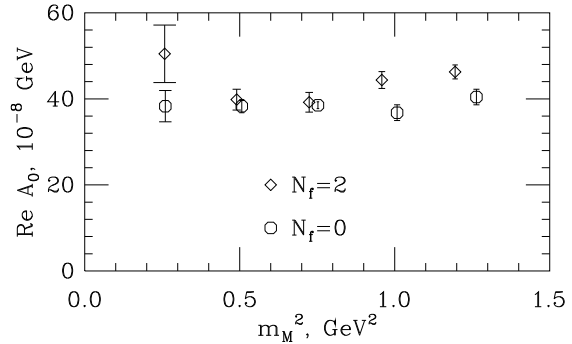}}
\vspace{-1.5cm}
\end{center}
\label{A0quench}
\caption{The effect of unquenching on $\Re A_0$. The lattice spacings
of quenched and dynamical ensembles are comparable.}
\vspace{-0.5cm}
\end{figure} 

%\section{$\Delta I=1/2$ rule}

\begin{figure}[htb]
\begin{center}
\leavevmode
%\vspace{-1cm}
\centerline{\epsfysize=5cm \epsfbox{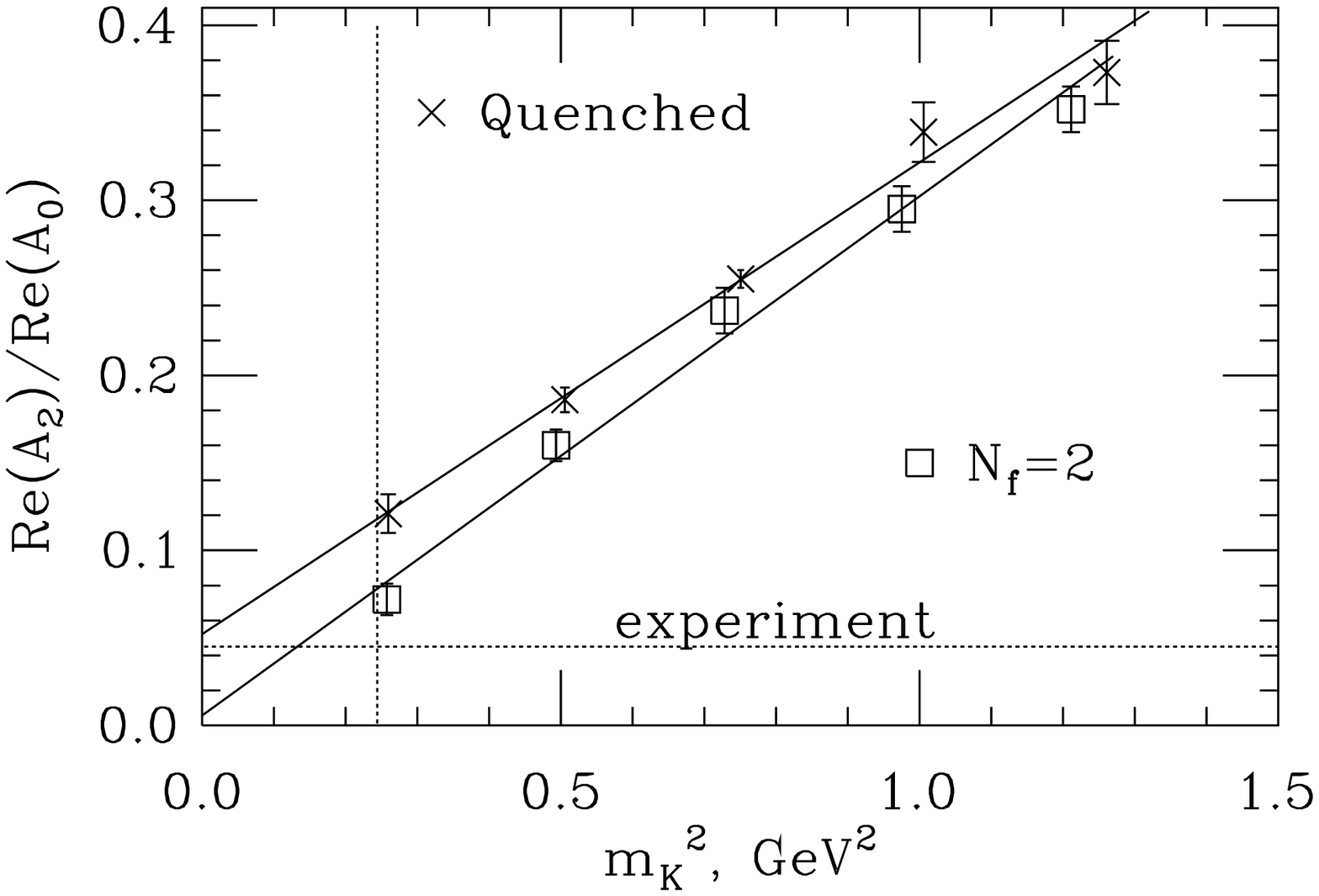}}
\vspace{-1.5cm}
\end{center}
\label{omega}
\caption{Isospin amplitude ratio vs. meson mass squared for quenched 
and dynamical ensembles with comparable lattice spacings. This enormous
dependence on the meson mass comes entirely from the behavior
of $\Re A_2$ (see Fig.~2).}
\vspace{-0.5cm}
\end{figure} 

We show the ratio $\Re A_2/\Re A_0$ for quenched 
($\beta=6.0$) and dynamical data sets in Fig.~7. The result is of the
same order of magnitude as experiment. However, the dependence on
the meson mass is so large that it prevents us from a more definite
conclusion. Clearly, we need higher order ChPT
terms to elucidate this subject further.

\section{Operator matching and \epsp  perspective}

Lattice and continuum operators have to be matched, for example 
by perturbation theory:

\begin{eqnarray}
\displaystyle
O_i^{\it cont}(q^*) = & O_i^{\it lat} + 
\displaystyle\frac{g^2(q^*a)}{16\pi^2}\displaystyle\sum_j(\gamma_{ij}
\ln (\frac{q^*a}{\pi} ) \nonumber \\
& + C_{ij})O_j^{\it lat} + O(g^4) + O(a^n) 
\end{eqnarray}

One-loop perturbative matching works well for operators $O_1$ and $O_2$
(relevant for $\Re A_0$): the corrections are small, and so is the $q^*$ 
dependence. However, for 
operators $O_5$ -- $O_8$ (relevant for \mbox{\epsp  )} the situation is 
much worse: several of the perturbative coefficients have not yet been
calculated, while others (notably $C_{PP}$) are too large at one-loop
order to trust the perturbation theory. Thus, a non-perturbative 
matching procedure is necessary to calculate \epsp .

%We have not performed this procedure thoroughly, but instead
%we came up with a temporary ``fix'', which can be viewed as a first step.
%We notice that at one-loop level most of the renormalization of $O_6$ 
%comes from corrections to each bilinear. If we knew the
%bilinear renormalization reliably, the rest of $O_6$ renormalization
%at one-loop level is small, so it seems reasonable that the ``improved''
%perturbation theory will work better, just as in the case of
%tadpole improvement. Eventually, this assumption needs to be checked by a 
%full non-perturbative study. 

\begin{figure}[!t]
\begin{center}
\leavevmode
\centerline{\epsfysize=5cm \epsfbox{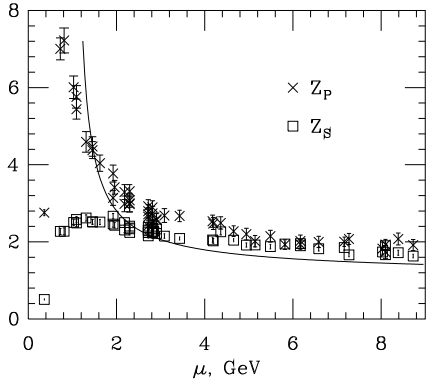}}
\vspace{-1.5cm}
\end{center}
\label{Zp}
\caption{Non-perturbative renormalization factors $Z_S$ and $Z_P$ 
in $\overline{\mbox{MS}}$ vs. momentum scale $\mu$ for $\beta=6.2$. 
The curve is the one-loop perturbation theory expectation (which does
not distinguish between $Z_S$ and $Z_P$ in the chiral limit).}
\vspace{-0.5cm}
\end{figure} 

As a first step towards this procedure we have computed bilinear 
renormalizaton factors $Z_S$ and $Z_P$ using a non-perturbative
method based on the strategy by Martinelli {\it et al.}~\cite{martinelli}
(see Fig.~8). Our best value for the strange quark mass in continuum
limit obtained with non-perturbative $Z_S$
in $\overline{\mbox{MS}}$ at 2 GeV is $103\pm 8$ MeV, which is
quite close to the one-loop result. 

$Z_P$ can be used to get a reasonable estimate of the renormalization
of operators $O_6$ and $O_8$. 
The value of $\langle O_6\rangle$
obtained in this way is very different from the tree-level value: it is
much smaller, close to zero.
This would produce a negative \epsp, contrary to experiment.
However, to give a more definite prediction we need to perform
a full non-perturbative renormalization procedure. 

%Somewhat unexpected is that after this ``semi-nonperturbative'' procedure
%$\langle O_6\rangle $ reverses its sign: it is positive and small. 
%This makes $\varepsilon '/\varepsilon$ negative and sizable, which seems
%to be in conflict with current experimental findings. There is a 
%remaining uncertainty about unknown $C_{ij}$ coefficients
%for operators $O_5$ and $O_7$, which may now play an important role 
%in \epsp  . In addition, the common uncertainties of final state interactions,
%chiral extrapolation and finite lattice spacing make a definite prediction
%for \epsp next to impossible. We hope to resolve some of these issues in the 
%future. 

\section{Summary}

We have obtained a reasonable statistical precision in studying
$\Re A_0$ and $\Re A_2$ as well as all matrix elements needed
for \epsp . The biggest uncertainties in our prediction
of the $\Delta I=1/2$ rule are higher order ChPT terms
and continuum extrapolation. In addition, a non-perturbative
operator matching needs to be done for \epsp . 

We thank Ohio Supercomputer Center and NERSC for CRAY-T3E time and Columbia
University group for their dynamical ensemble.

\end{document}